\documentstyle [epsf,here,rotate] {article}
\begin{document}

\centerline{\LARGE {\bf Gauge Lattice Simulation of the soft QGP Dynamics }} 
\centerline{\LARGE {\bf in Ultra-Relativistic Heavy-Ion Collisions  } }
\vspace{0.1cm} 
\centerline{ W. P\"oschl and B. M\"uller}
\vspace{0.1cm} 
\centerline{Department of Physics, Duke University, Durham, NC 27708-0305, USA}
\vspace{0.5cm} 
\vspace{0.0cm}
\abstract
{
\noindent
In a fully relativistic approach, a RLSM description of
nuclei colliding at ultra-relativistic energies can be
formulated within the framework of a classical transport theory.
The valence quarks of the nucleons 
are described through collections of classical point-like particles 
moving in the continuum. They are coupled to soft gluon fields 
which are described through the Yang Mills equations on a gauge lattice. 
In a first step, we focus on the range of low-$p_t$ interactions.
Results from numerical model simulations of pure gluonic
nucleus-nucleus collisions on SU(2) gauge lattices in 3+1 dimensions 
are presented. They show an effect which we call the glue burst.
\hfill\break
}

\vspace{1.0cm}
\noindent
The state of super-dense nuclear matter called the quark gluon plasma
is expected to occur in the central kinematic region of ultra-relativistic
heavy-ion collisions. 
In view of upcoming experiments at the RHIC in 
Brookhaven and the LHC at CERN,  
it is one of the most challenging topics in the physics of
ultra-relativistic heavy-ion collisions to develop a coherent 
description of the formation of the quark gluon plasma state.
Many of the descriptions which have been developed over the past
years are based on the idea of a perturbative scattering
of partons within transport models. One of the problems in these
descriptions concerns the initial state of the colliding nuclei.
The transport equations start from probability distributions of
partons in the phase space. In reality however, the states of
the nuclei are described by coherent parton wave functions.  
The incoherent parton description becomes inadequate at exchanges
of small transverse momenta.
A four years ago McLerran and Venugopalan proposed 
\cite{McLerran.94} that the proper 
solution of these difficulties is the perturbative expansion not around
the empty QCD vacuum but around a vacuum of the mean color fields
which accompany the quarks in the colliding nuclei. This idea motivates
a combination of the parton cascade model \cite{Geiger.92} with
a mean field description of the color fields.
Before this can be done successfully, it is important to study the
non-perturbative dynamics of the color mean-fields themselves.

\noindent
In Fig. 1, a typical scenario
of a collision is displayed for times shortly before and after the collision.
The figure depicts the basic idea behind our model of a combined 
particle-gauge lattice system. The collision describes a 
scenario in which two media are involved: The valence quarks, described
through collections of point-like color charged particles (solid lines)
and the gluon fields (helical lines). While the quarks interact essentially
only through the exchange of gluons, the gluon fields are self-coupled.
%
%
%
%
%
%
\begin{figure}[H]
\centerline{
\epsfysize=8cm \epsfxsize=8cm
\epsffile{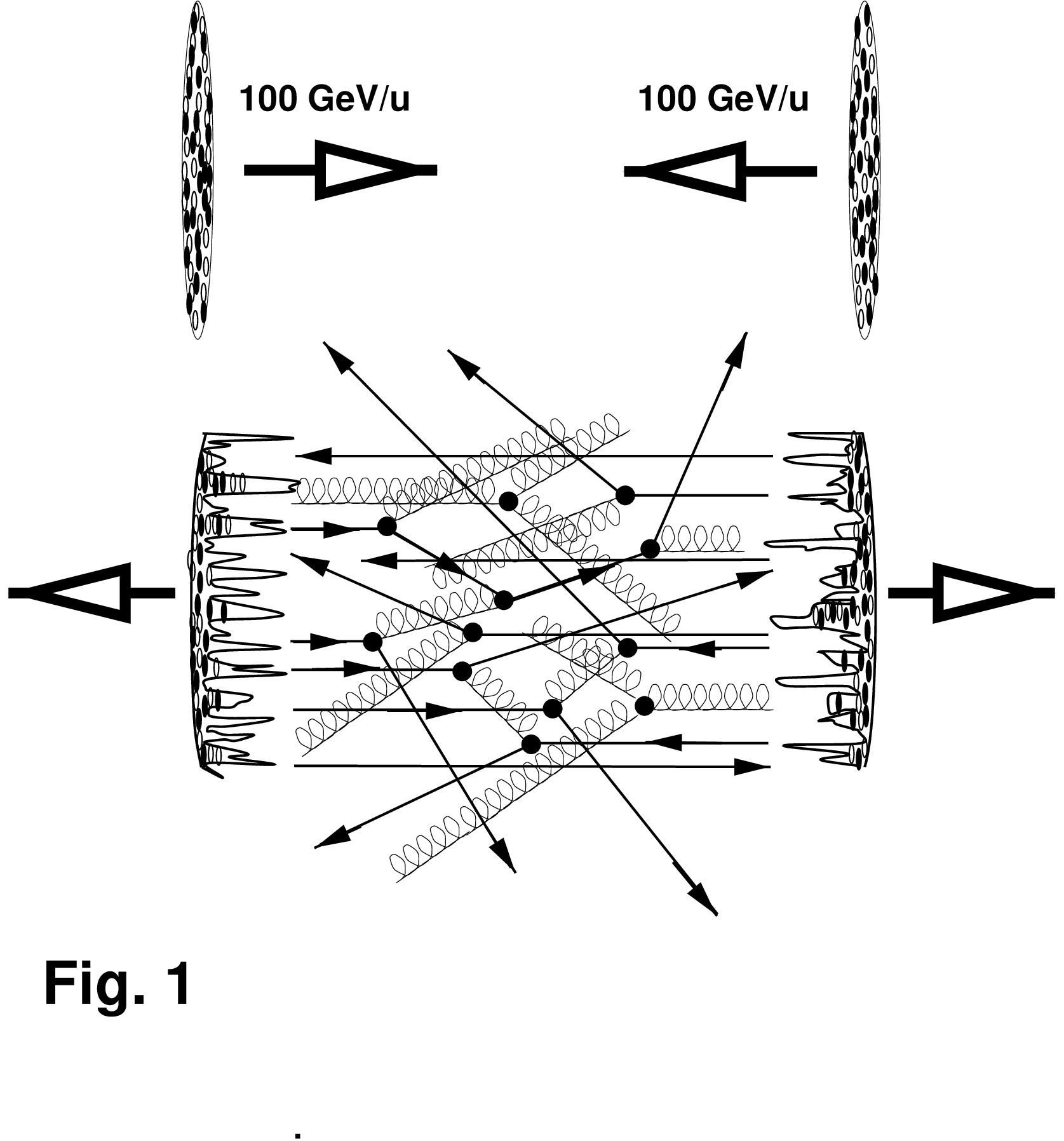}
\hspace{0.5cm}
\epsfysize=8cm \epsfxsize=8cm
\epsffile{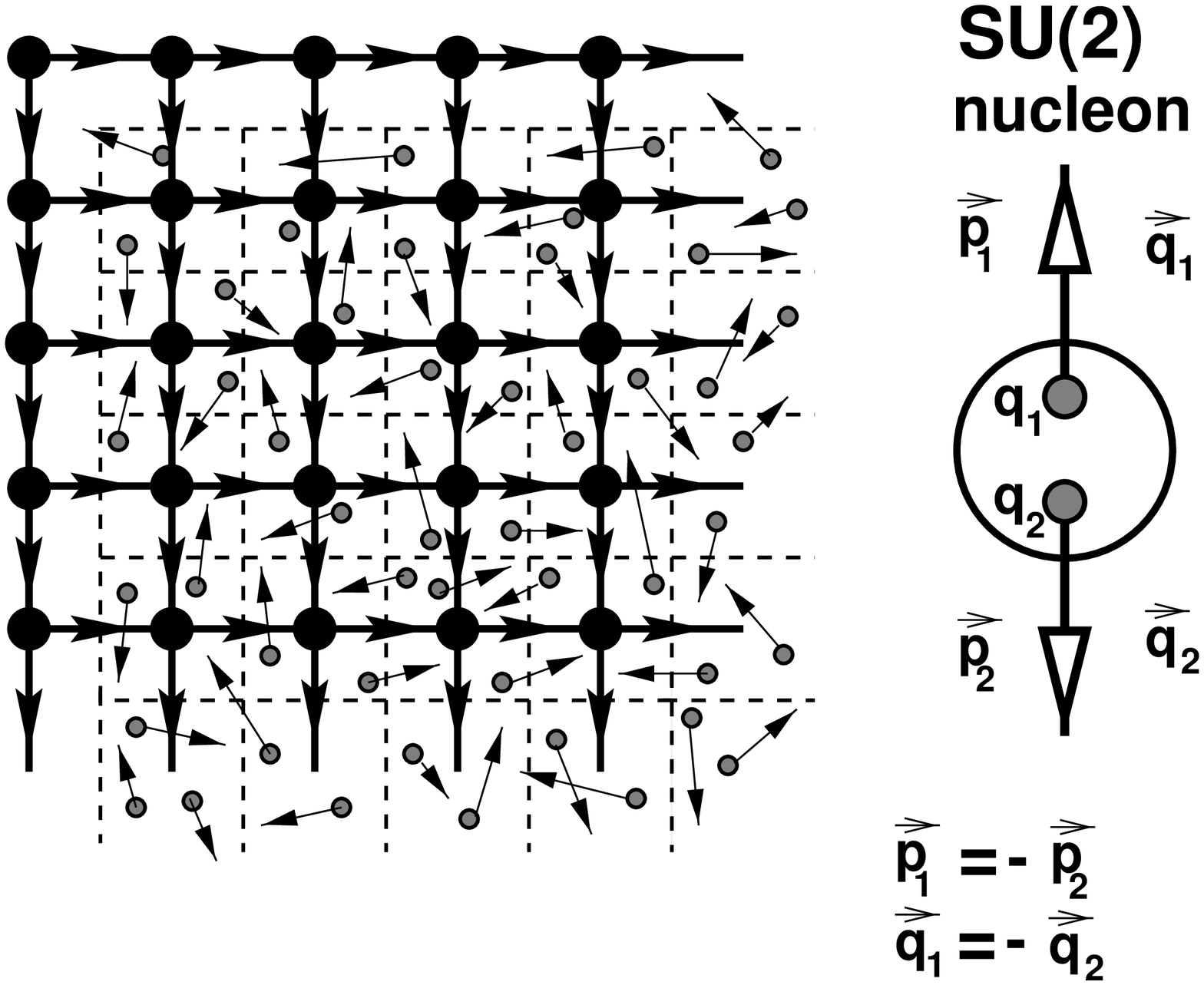}
}
\end{figure}
\noindent
The right-hand part of Fig. 1 displays the basic concept 
of our gauge lattice transport model \cite{BMP.98}.
The soft color fields are represented on a SU(2) gauge
lattice in three dimensional Euclidean space and are described by
the Yang-Mills equations. The quarks are
represented as collections of color charged particles moving
in the continuous background in position space. Their trajectories
are solutions of the Wong equations. The particles
are coupled with the lattice by energy and color exchange mechanisms.
In SU(2) a nucleon initially 
is represented through a pair of quarks with opposite momenta and
opposite color charge. In principle the model allows
the inclusion of collision terms to describe the hard collisions
between the particles. This idea has recently been submitted
\cite{BMP.98}. 

\noindent
In recent simulations of a central collision of two
nuclei with the above described model \cite{PM.98}    
it turned out that in addition to large transverse
gluon radiation there exists a contribution to the
transverse dynamics from ``soft glue field scattering''.
This leads to the question, how important is really the mean field 
dynamics for the transverse dynamics. Due to non-linear effects,
the gluon mean field might contribute more to the
transverse momenta than naively expected. We, therefore, go first
one step back and focus on the pure gluon mean field dynamics
in the non-perturbative regime and leave out the particles.

\noindent
In short denotation the homogeneous Yang-Mills equations read
${\cal D}_{\mu}{\cal F}^{\mu\nu} = 0$.
An lattice version of the continuum Yang Mills equations 
is constructed by expressing the field amplitudes as elements of the 
corresponding Lie algebra, i.e. ${\cal E}_{{\vec x},k},
{\cal B}_{{\vec x},k}\,\in$ LSU(2) at each lattice site $\vec x$.  
We choose the temporal gauge $A_0 = 0$ and define the following variables. 
\begin{eqnarray}
\label{Eq.6}
{\cal U}_{{\vec x},l} 
= \exp(-iga_l{\cal A}_l(x)) \,\,=\,\, 
{\cal U}^{\dagger}_{{\vec x}+l,-l} \qquad
{\cal U}_{{\vec x},kl} 
= {\cal U}_{{\vec x},k}\, {\cal U}_{{\vec x}+k,l}\, 
    {\cal U}_{{\vec x}+k+l,-k}\, {\cal U}_{{\vec x}+l,-l} 
\end{eqnarray}
In adjoint representation the color electric and color magnetic
fields are expressed in terms of the above defined 
link variables ${\cal U}_{{\vec x},l}$ and plaquette variables
${\cal U}_{{\vec x},kl}$ in the following way.
\begin{eqnarray}
\label{Eq.9}
{\cal E}_{{\vec x},j} = { {1}\over{iga_j}}\,\dot {\cal U}_{{\vec x},j}
                   {\cal U}^{\dagger}_{{\vec x},j}    \qquad
{\cal B}_{{\vec x},j} = { {i}\over{4ga_ka_l}}\,\epsilon_{jkl}\,
\bigl({\cal U}_{{\vec x},kl} - 
{\cal U}^{\dagger}_{{\vec x},kl} \bigr).
\end{eqnarray}
The lattice constant in the spatial direction $l$ is denoted by $a_l$.  
We choose ${\cal U}_{{\vec x},i}$ and ${\cal E}_{{\vec x},i}$ as the 
basic dynamic field variables and
numerically solve the following equations of motion.
\begin{eqnarray}
\label{Eq.12}
\dot{\cal U}_{{\vec x},k}(t) &=& i\,g\,a_k\,{\cal E}_{{\vec x},k}(t)\,{\cal U}_{{\vec x},k}(t) 
\qquad\qquad\qquad\qquad\qquad \\
\label{Eq.13}
\dot{\cal E}_{{\vec x},k}(t) &=& {i\over{2ga_1a_2a_3}} \sum\limits^3_{l=1}
\Bigl\{{\cal U}_{{\vec x},kl}(t) -
{\cal U}^{\dagger}_{{\vec x},kl}(t) 
\,\Bigr.
-  \Bigl. {\cal U}^{\dagger}_{{\vec x}-l,l}(t)\,
\Bigl( {\cal U}_{{\vec x}-l,kl}(t) -
{\cal U}^{\dagger}_{{\vec x}-l,kl}(t)\Bigr)\,
{\cal U}_{{\vec x}-l,l}(t)\, \Bigr\}.
\end{eqnarray}

\noindent
We first study the collision of plane wave packets which have
initially constant amplitude in the planes transverse to the wave vector.
We choose periodic boundary conditions implying the topology of a 3-torus
for our lattice.
The size is chosen as 10x10x2000 lattice points. 
For a collision, we initially arrange two Gaussian wave packets
with average momenta $(0,0,\pm\overline k_z)$ and width
$\Delta k_l$ between the maximal and minimal momenta
on the lattice:
$ k_l^{min}\,\ll\, \Delta k_l\, \ll\, \overline k_l\,\ll\, k_l^{max} $
The polarization in color space is defined through the unit
vectors $\vec n_L^c$ for the left and $\vec n_R^c$ for the left (L) 
and right (R) moving wave packet, respectively. 

\noindent
We express the initial conditions through the gauge fields \cite{Hu.95}
\begin{equation}
\label{Eq.20}
A_{R/L}^c=\delta_{l1}{\vec n}^c_{L/R}\phi(\mp t,z-Z_{R/L})
\end{equation}
where the scalar factor $\phi(t,z)$ defines the initial wave packet
\begin{eqnarray}
\label{Eq.21}
\phi(t,x_3)&:=&\phi_0
\exp{\bigl(-{1\over 2}\Delta k_3^2(t+x_3)^2\bigr)}
\cos{\bigl(\overline k_3(t+x_3)\bigr)}.
\end{eqnarray}
The amplitude factor is defined as
$ \phi_0 :=
\sqrt{ {(2\Delta k_3)}/({\sqrt{\pi}\sigma\overline k_3})} $
where the parameter $\sigma$ denotes the transverse area per quantum 
contained in the wave field.
We choose the polarizations in color space 
$\vec n_R=\vec n_L = (0,0,1)$ for the
Abelian case and $\vec n_R = (0,0,1),\,\vec n_L = (0,1,0)$
for the non-Abelian case.
Once the initial fields are
mapped on the lattice, the time evolution of the collision
starts from the superposed initial conditions
\begin{equation}
\label{Eq.24b}
{\cal U}^{(0)}_{{\vec x},l}=
{\cal U}^{(0)}_{{\vec x},l,L}\cdot{\cal U}^{(0)}_{{\vec x},l,R},
\qquad
{\cal E}^{(0)}_{{\vec x},l}=
{\cal E}^{(0)}_{{\vec x},l,R}+{\cal E}^{(0)}_{{\vec x},l,L},
\end{equation}
respectively.
A linear superposition of solutions obeys the Yang-Mills 
equations only if these solutions have overlap zero. Therefore,
the initial separation $\Delta Z$ of the wave packets may not be chosen
to small.

\noindent
The calculation contains four parameters.
The relative isospin polarization which is parameterized through
the angle $\theta_C$. The average momentum of the wave packets
$\overline k$ and their width $\Delta k_3$ and the coupling 
constant g which can be rewritten in terms of the parameter
$\sigma$ as $g'=g/\sqrt{\sigma}$ by 
simultaneously rescaling the field ${\cal A}$ as 
${\cal A}'=\sqrt{\sigma}{\cal A}$.
Consequently, the system shows the same dynamics for different
values of g and $\sigma$, as long as the ratio 
$g'=g/\sqrt{\sigma}$ is kept fixed.

\noindent
In the following we present results from a simulated collision in the
non-Abelian case with the parameters 
$\overline k_3=\pi/(2a_3)$,
$\Delta k_3=\pi/(100 a_3)$, $g=1$, $\sigma = 1$.
Further, we used $a_l = 0.1\,{\rm fm}$ and $\Delta t= 0.01\,{\rm fm}$.
Subsequently, we refer to the direction of the collision
axis (the z-axis) as the ''longitudinal direction'' and to directions
perpendicular to the collision axis as the ''transverse directions''.
Accordingly, we define the transverse and longitudinal energy densities
of the color electric field 
and the transverse and longitudinal energy flow distributions
\begin{equation}
\label{Eq.25a}
w_T^{(E)}(t,z)=\sum\limits_{x,y} a_1a_2 \sum\limits_{l=1}^2 
                 Tr\bigl({\cal E}_{{\vec x},l}{\cal E}_{{\vec x},l}\bigr),
\qquad
w_L^{(E)}(t,z)=\sum\limits_{x,y} a_1a_2 
                Tr\bigl({\cal E}_{{\vec x},3}{\cal E}_{{\vec x},3}\bigr)
\end{equation}
Fig. 2 displays $w_T^{(E)}(t_n,z)$ plotted over the z-coordinate 
at various time steps $t_n$ as indicated on top of the curves.  
%
%
%
%
%
%
\begin{figure}[H]
\centerline{
\epsfysize=7cm \epsfxsize=8cm
\epsffile{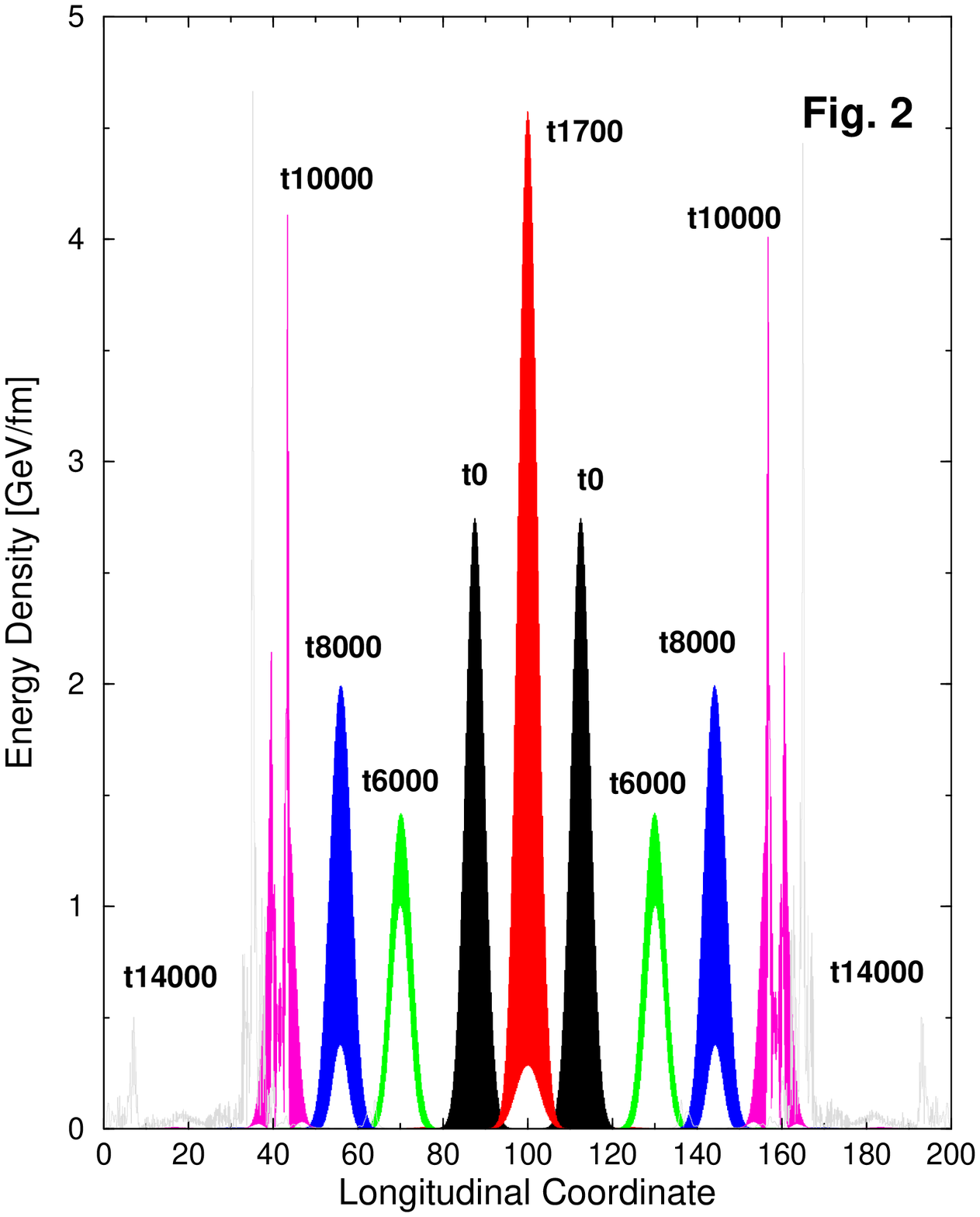}
\hspace{1.0cm}
\epsfysize=7cm \epsfxsize=8cm
\epsffile{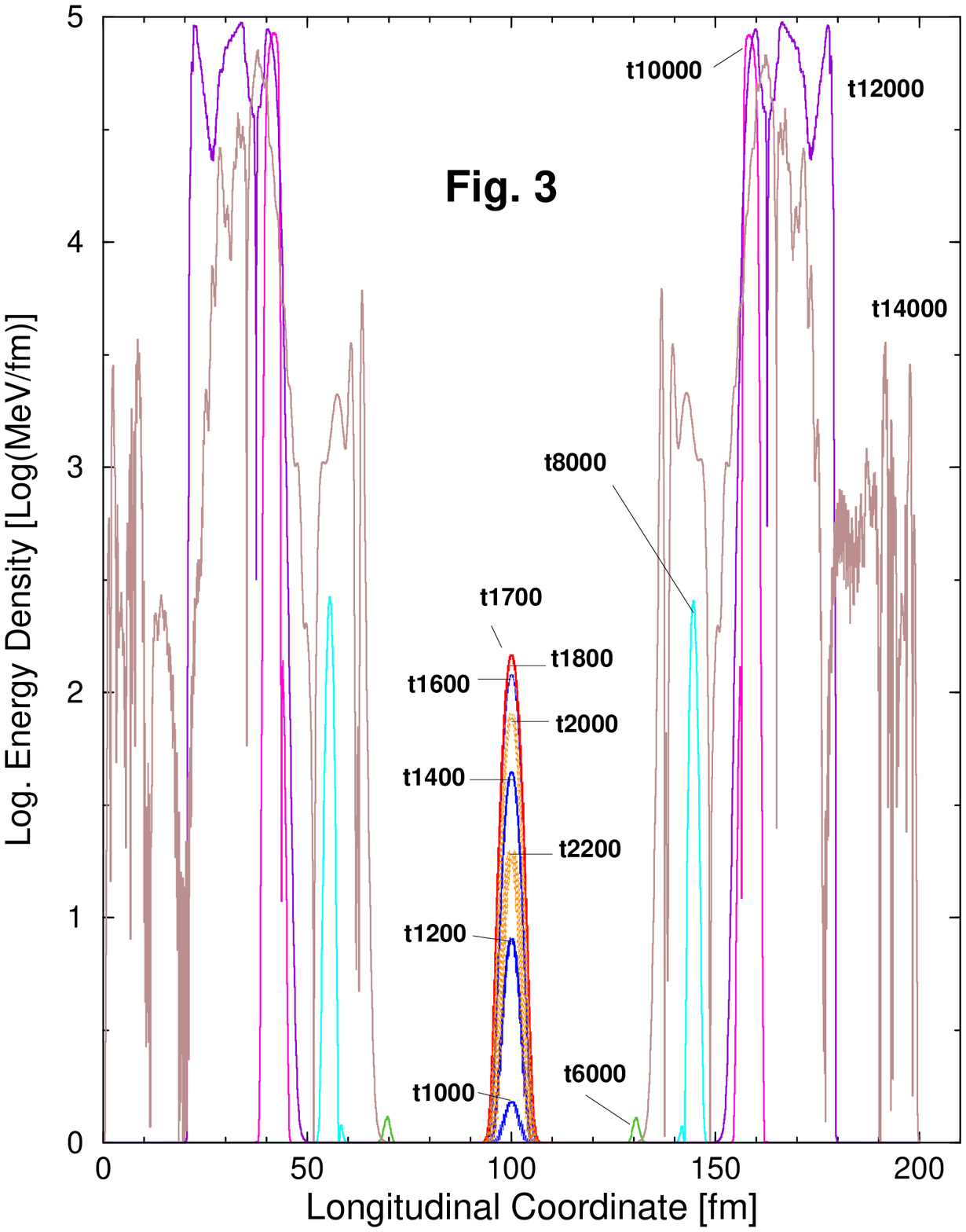}
}
\caption{Transverse energy densities $w_T(t,x_3)$ are
         displayed for selected time steps $t_n$ }
\end{figure}
\noindent
At initial time $t_0$ the curves $w_T^{(E)}(t_0,z)$ of the wave packets 
are completely filled indicating the strong
oscillations according to $\overline k_3=\pi/(2a_3)$.
After 1700 time steps the wave packets are colliding and have reached
maximum overlap. The smooth white zone at the bottom indicates the
appearance of long wavelength modes in the Fourier spectrum of the
common field. These low frequent modes have been excited in the
penetration time. At time step $t_{6000}$, almost all the
energy which has originally been carried by short wavelength modes
around $\overline k_3$ is transmitted into long wavelength modes.
This can be seen from the high frequent oscillations on the surface
of the big white zones in the two receding humps. At time step
$t_{8000}$ however it seems that energy is transmitted back into
high frequent modes and finally the wave packets start to decay
around time step $t_{10000}$.   
Simultaneously, we display the corresponding longitudinal
energy densities $w_L^{(E)}(t,z)$ in Fig. 3 for the same
time steps $t_n$ but on a logarithmic scale. 

\noindent
We remember that the wave packets were
initially polarized into the transverse $x$-direction and
consequently $w_L^{(E)}$ has to be zero as long as
the wave packets propagate freely. However, when the two
colliding wave packets of different color start overlapping,
we observe an increasing longitudinal energy density in
the overlap region around the center of collision at
$z=100\,{\rm fm}$. Fig. 3 clearly shows that $w_L^{(E)}(t,z)$
grows rather fast from time step $t_{1000}$ until time step
$t_{1700}$ where it reaches a maximum and has grown 
exponentially by more than two orders of magnitude. 
For larger times the hump decreases and practically 
disappears at $t_{3000}$. Around $t_{6000}$, however, the longitudinal
energy density grows again at the positions of the receding
wave packets. After 10000 time steps $w_L^{(E)}(z)$ has
increased by five orders of magnitude. 
Now the question arises whether the
energy deposit in longitudinal links can be associated with
fields propagating into transverse directions. 
The transverse and longitudinal energy flow densities displayed
in Fig. 4 and Fig. 5 are defined by the Poynting vector 
\begin{eqnarray}
\label{Eq.27}
{\vec {\cal S}} = c\,\vec{\cal E}\times \vec {\cal B},
\qquad
s_T(t,z):= 2\sum\limits_{l=1}^2\sum\limits_{x,y} a_1a_2
\big\vert \mbox{\rm Tr}\bigr( \mbox{\cal S}_l(t,x,y,z)\bigr)\big\vert,
\qquad
s_L(t,z):= 2\sum\limits_{x,y} a_1a_2
\big\vert \mbox{\rm Tr} {\cal S}_3(t,x,y,z)\big\vert.
\end{eqnarray}
%
%
%
%
%
\begin{figure}[H]
\centerline{
\epsfysize=7cm \epsfxsize=8cm
\epsffile{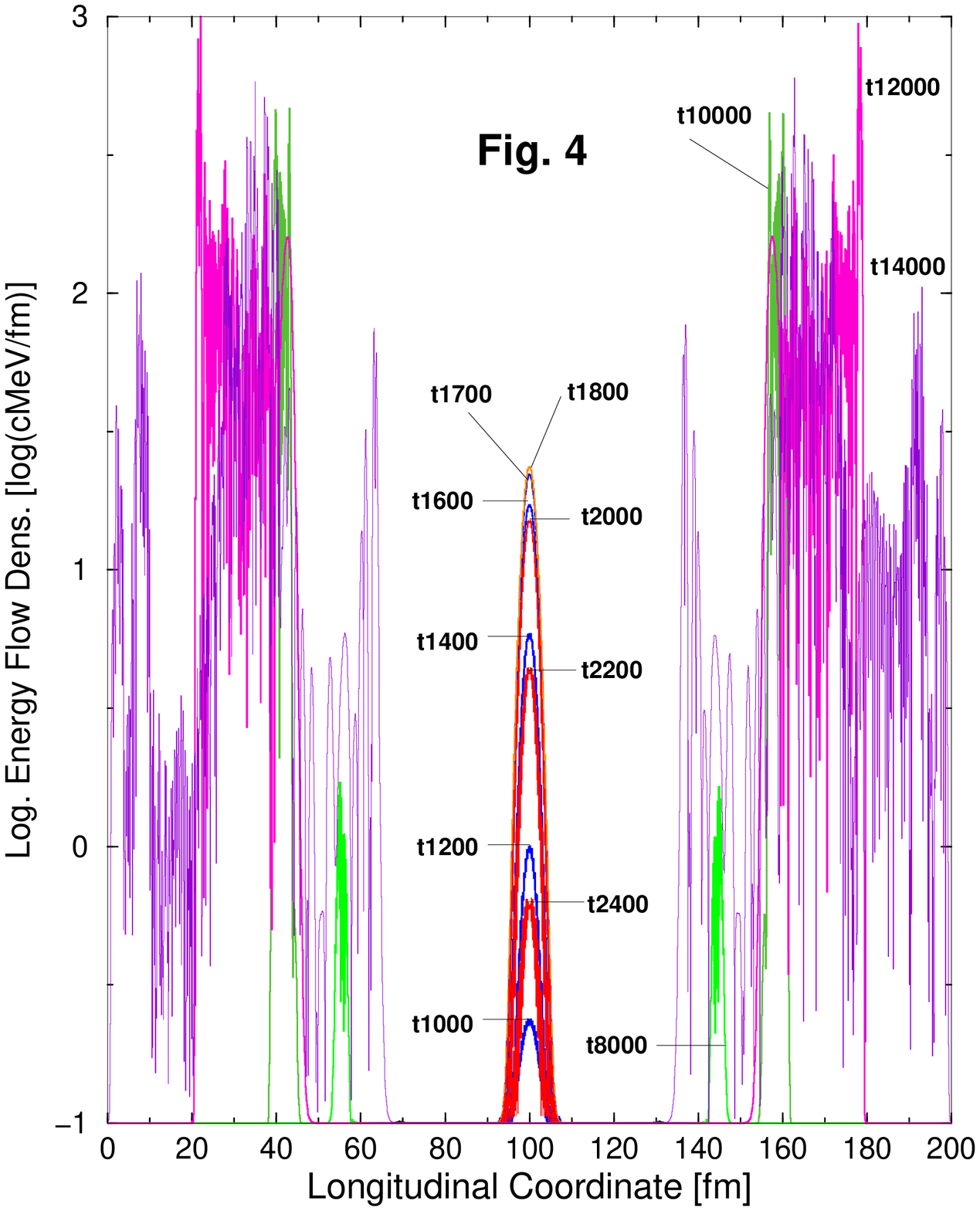}
\hspace{1.0cm}
\epsfysize=7cm \epsfxsize=8cm
\epsffile{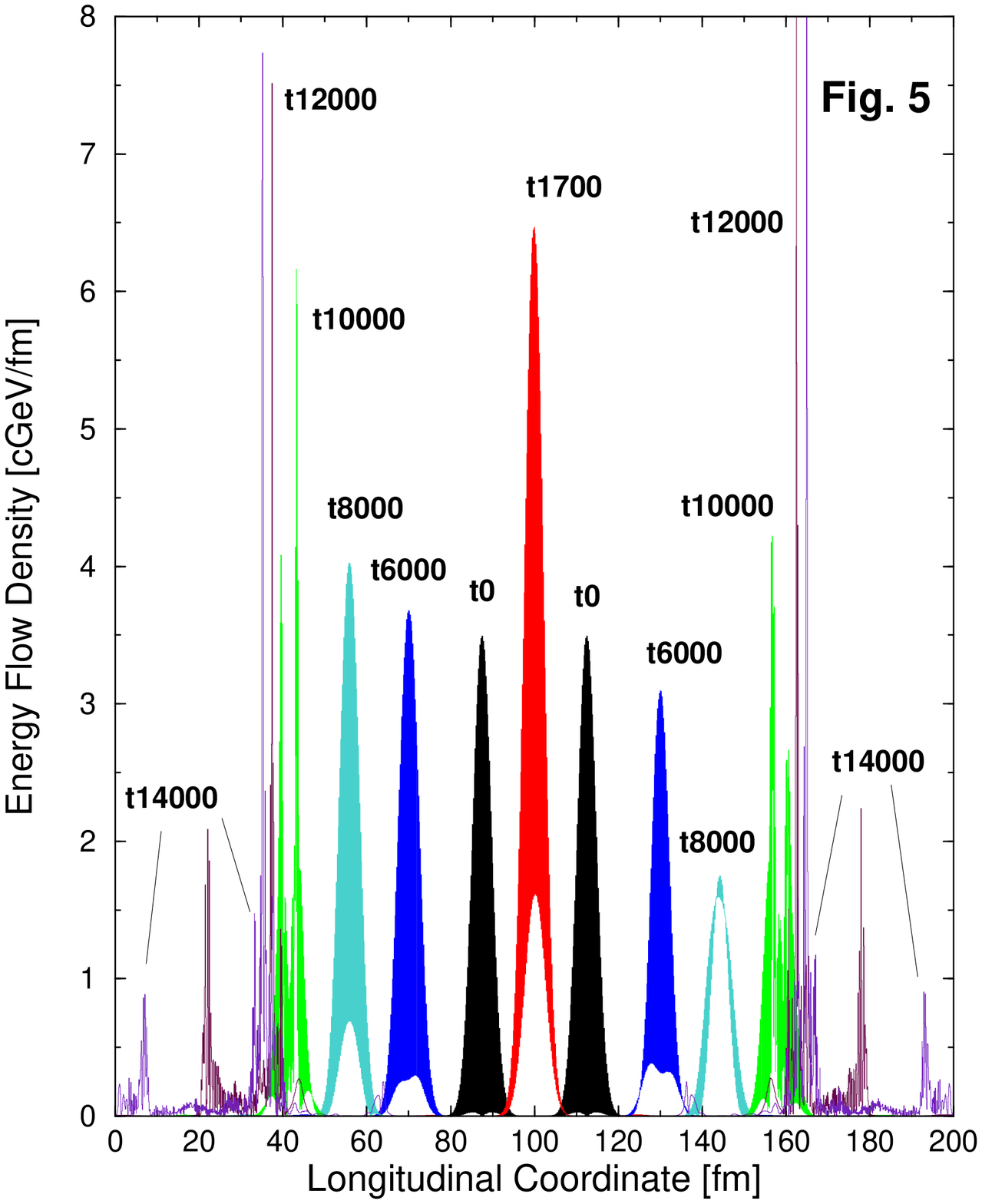}
}
\end{figure}
\noindent
A direct comparison of Fig. 4 with Fig. 3 shows that 
the transverse energy flow distribution $s_T(t,z)$ 
behaves very similar as compared
to the longitudinal energy distributions. We call
the phenomenon resulting in 
the strong increase of the transverse energy flow at large
times the ''glue burst''. 
The most interesting feature 
is the time delay extending from time step 
$t_{1600}$ up to $t_{6000}$. Repeating the calculation for increasing g-values,
we found that it scales with $g^{-2}$. 
The explanation can be found through a carefull analysis of the 
Yang-Mills equations. 
The growing peak in the center of collision
clearly shows that the wave packets interact strongly due
to the non-linearity of the Yang Mills equations.
The wave packets, however, are restored after they have
passed through each other and look for a while almost 
unchanged. Suddenly, around time step $t_{6000}$ they start to decay.
At time step $t_{10000}$ the transverse energy flow density has grown
by three orders of magnitude in comparison to the one at time
step $t_{6000}$. Afterwards the energy flow density shows no further
increase. A calculation for the Abelian case with
same parameters shows no fragmentation of the wave packets, no
transverse energy currents
and no peak-like structure in the overlap region.    

\noindent
Fig. 5 displays the corresponding longitudinal energy current
densities $s_L(t,z)$. As a function of time they behave 
similarly as the transverse energy densities
$w_T^{(E)}(t,z)$ shown in Fig. 2. The graphs in Fig. 4 do not appear
to be completely symmetrical. Matinyan et al. have shown,
that the Yang Mills equations are sensitive
to small perturbations. The difference in shape of the
right moving and left moving energy flow distribution
reveals a small difference in the initial conditions by reading
them slightly asymmetrically on the lattice. 
This effect is enhanced in our example
because we have chosen a very short initial wavelength 
$\overline k_3=\pi/(2a_3)$. At high oscillations in position space, 
slight phase shifts can strongly change the shape of the density distribution. 
Nevertheless, the total integrated energies of both wave packets agree
exactly. Fig. 2 to Fig. 5 are created from data obtained in one run of
our computer code. A comparison of Fig. 5 with Fig. 2 shows that
the asymmetry comes from the B-fields which seem to be more sensitive
to perturbations. Further, a comparison of Fig. 4 with Fig. 5 shows
that $s_L(t,z)$ in Fig. 5 decreases when $s_T(t,x_3)$ in Fig. 4 increases.  
At time step $t_{14000}$ a sizable fraction of the longitudinal energy flow is
converted into transverse energy flow. 
We have repeated the calculations at
$\overline k_3 = \pi/(100a_3)$ and found qualitatively
the same results. Here, however we demonstrate that our results
are even found at the critical average longitudinal wavelength of $4a_3$
which is just twice as long as the shortest wavelength of $2a_2$ at
the lattice cut-off. We also observe that the transverse energy current
increases for larger $\overline k_3$.  

\noindent
In a study of colliding wave packets of finite extension into one 
transverse direction we use the same initial conditions as in the
case of plane waves but multiply the function $\phi$ in Eq. (\ref{Eq.21})
with a factor $\exp{\bigl(-(\Delta k_y y)^{\lambda})}$. 
In the simulation of the collision shown in Fig. 6 below we used the
parameter values $\Delta k_y = 1/(40a_2)$ and $\lambda=8$ on a
4x200x800 lattice. The coupling constant was taken $g=6.0$ in order to
reduce the time delay. The large value of $\lambda$ leads to a width of
$8\,{\rm fm}$ in the y-direction leaving a space of $6\,{\rm fm}$ on each side
of the initial wave packet. Similar results as for colliding plane waves
were found. The induced transverse energy currents, however, were
much larger. Fig. 6 shows the transverse energy distribution at the
time steps $t100$, $t1500$, $t3500$, $t4000$, $t5000$, and $t6000$.
The glue burst occurs around time step $t4000$. The full explanation
of this phenomenon will be published elsewhere. \hfill\break

\noindent
Note: The Fig. 6 exceeds the size accepted at the lanl-server and
was therefore omitted. We recommend to download the postscript file
from the web page http://www.phy.duke.edu/\~\,poeschl/ under 
``Rsearch Related Links'' and ``Preprints''.
%
%
%
%
%
%
%
%
%
%
%


\end{document}